# Density Functional Theory Study of the Hydrogen Evolution Reaction in Haeckelite Boron Nitride Quantum Dots

Rupali Jindal[1], Vaishali Sharma[1], and Alok Shukla[1,*]

[1]*Department of Physics, Indian Institute of Technology Bombay, Powai, Mumbai 400076, India*

*Corresponding author email: shukla@phy.iitb.ac.in

**Abstract**

To satisfy arising energy needs and to handle the forthcoming worldwide climate transformation, the major research attention has been drawn to environmentally friendly, renewable and abundant energy resources. Hydrogen plays an ideal and significant role is such resources, due to its non-carbon based energy and production through clean energy. In this work, we have explored catalytic activity of a newly predicted haeckelite boron nitride quantum dot (haeck-BNQD), constructed from the infinite BN sheet, for its utilization in hydrogen production. Density functional theory calculations are employed to investigate geometry optimization, electronic and adsorption mechanism of haeck-BNQD using Gaussian16 package, employing the hybrid B3LYP and wB97XD functionals, along with 6-31G(d,p) basis set. A number of physical quantities such as HOMO/LUMO energies, density of states, hydrogen atom adsorption energies, Mulliken populations, Gibbs free energy, work functions, overpotentials, etc., have been computed and analysed in the context of the catalytic performance of haeck-BNQD for the hydrogen-evolution reaction (HER). Based on our calculations, we predict that the best catalytic performance will be obtained for H adsorption on top of the squares or the octagons of haeck-BNQD.We hope that our prediction of most active catalytic sites on haeck-BNQD for HER will be put to test in future experiments.

**Keywords:** Hydrogen evolution reaction; haeckelite boron nitride; quantum dots; density functional theory; Gibbs free energy

# 1. Introduction

Hydrogen as a clean, powerful, and efficient energy source gains sufficient attention because of climatic conditions, limitations of conventional resources, and increasing energy demands [1-4]. Hydrogen economy seems to be the best possible solution to the upcoming energy crisis and hazardous environment problems [5-8]. Presently, 96% of the hydrogen is produced through coal, oil, or natural gas, and only 4% carbon-free hydrogen production is carried out by splitting of water because of high cost [9]. Therefore, to make hydrogen a viable source of energy, we need to make its production as cost efficient as hydrocarbon-based fuels. Mass hydrogen production, purification, storage, and transportation are the factors one needs to consider for practical green hydrogen energetics [10-12]. In the electrochemical water splitting, a highly efficient catalyst is required to optimize the hydrogen evolution reaction (HER→ $2H^+ + 2e^- = H_2$) [13-14]. Platinum (Pt) is the most promising and widely used catalyst, however due to its scarcity and high cost, one needs to replace it, and other expensive metal catalysts, with cheap, high-performance ones to make hydrogen generation cost effective [15-19].

The research field of nanosheets is growing exponentially because of their unique properties and wide range of applications. One of the most widely studied two-dimensional (2D) nanomaterial, graphene, and its one-dimensional counterparts carbon nanotubes, have a large number of potential applications, including hydrogen storage and splitting. After these carbon-based materials, the 2D boron nitride nanotubes and nanosheets also have a vast research literature related to hydrogen adsorption and storage because of their immense chemical and thermal stability [20-23]. In the 2D nanosheets of BN, hexagonal boron nitride (h-BN) and recently detected haeckelite boron nitride (haeck-BN) are the most studied ones analogous to the carbon structures of graphene and octagraphene [24-31]. The haeckelite (haeck-BN) nanosheets have quite different structural and electronic properties as compared to their h-BN counterpart, because of the presence of the octagon and square rings, in addition to the hexagons [30]. These differences are reflected in bond lengths, bond angles, HOMO-LUMO gaps, work function etc. The band gap in haeck-BN is higher as compared to h-BN, but the work function reduces while going from h-BN to haeck-BN [30].Interestingly, the investigations of BN in the context of hydrogen storage have also been extended to hydrogen bubbles on hexagonal BN flakes [31]. While reducing the dimensionality of nanostructures to the zero-dimensional quantum dots, it is seen that besides developing useful physical, electronic and optical properties, quantum dots are also emerging as promising candidate in energy storage and conversion sector as electrocatalysts [32-34]. Significant attempts have been made to analyse the catalytic properties of quantum dots (QDs) in the context of HER [35-38]. The recent works on the graphene quantum dots (GQDs) in hydrogen production have generated considerable interest in the field after successful predictions for HER. Like GQDs, further investigations are in progress to find other thermally stable, non-toxic, inert chemical catalysts. Doping GQDs with B and N enhances their catalytic activity considerably [39,40]. Considering the effect of B-N doping on the electrocatalytic activity of QDs and the significant role of B-N nanostructures for hydrogenadsorption and storage, boron nitride quantum dots (BNQDs) appear to be promising substitutes because of their tailorable properties. BNQDs display confinement and edge effects, and allow us to tune electrical, optical characteristics by varying their shapes and sizes. This unique behaviour leads to many astonishing optoelectronic, chemical, and biochemical properties and practical applications in catalysis, lighting, and bioimaging [41-43].In present study, a new haeckelite structure of BNQD has been predicted, showing quantum confinement effects along with its catalytic activity for HER. So far, only the haeckelite structure of the GQD has been investigated to the best of our knowledge[44].

The first-principles density functional theory (DFT) has been employed to examine the electrical and structural properties of this novel haeckelite structure of BNQD in detail. Furthermore, the total density of states, overpotential, dipole moment, work function etc. have been analysed thoroughly to predict H adsorption on haeck-BNQD. The HOMO-LUMO gap of haeck-BNQD has been compared with the haeck-BN structure to gain insights into its confinement effects. The results of calculations of adsorption energies and Gibbs free energies have demonstrated favourable outcomes for hydrogen adsorption, and we have also found that haeck-BNQD can be a promising electrocatalyst for HER.

## 2. Computational Methodology

All the DFT calculations including total energies, geometry optimization, and vibrational frequency analysis have been performed using the computational chemistry software Gaussian16 [45]. For the visualization of structures Gaussview6.0 [46] was utilized, while the total density of states (TDOS) plots were generated using Multiwfn [47]. Closed-shell DFT calculations are performed on the pristine and H atom absorbed structures of haeckelite BNQDs using the B3LYP and the wB97XD exchange-correlation functionals. B3LYP is a hybrid functional composed of Becke's three-parameter exchange functional with Lee, yang and Parr correlation functional containing both local and non-local terms, and wB97XD is a long-range correlation functional that includes empirical dispersion and long range correlations. B3LYP functional is effective in predicting properties related to hydrogen adsorption and storage on various materials as reported in the literature [48-54]. While dealing with systems involving ad-atoms, van der Waals interactions play a significant role. To account for them, we have used the wB97XD functional that includes the London dispersion corrections arising due to long-range interactions. In the wB97XD functional, the attractive van der Waals correction term of $-C_6/R^6$ was included by Grimme, also known as the GD2 dispersion model. Mathematically, the dispersion correction term is given by:

$$E_{disp}^{D2} = -S_6 \sum_{i,j>i}^{N_{atoms}} \frac{C_6^{ij}}{(R_{ij})^6} f_{damp}(R_{ij}) \qquad (1)$$

where $f_{damp}(R_{ij}) = \frac{1}{1+a(R_{ij}/R_r)^{-12}}$ is the damping function. Here, $R_r$ is the sum of vdW radii of the atomic pair *ij*, and the only non-linear parameter, *a*, controls the strength of dispersion corrections. $N_{atoms}$ in Eqn. (1) shows number of atoms in the system. $C_6^{ij}$ is the dispersion coefficient for atom pair *ij*. $R_{ij}$ presents interatomic distance. The correlational functional dependence of this additive dispersion term was included by the fitting parameter s6 and sr,6 used in the damping function. The values for these two parameters for the wB97XD functional are 1.0 and 1.1 respectively. The split valence 6-31G basis set along with d,p polarization function was used for all the calculations.

In the first step, the geometry of the haeck-BNQD is optimized using aforementioned functionals. Next, the geometry optimization is performed with H adsorbed haeck-BNQD systems. In order to ensure the stability of the optimized structures, we computed their vibrational frequencies. Real values of the frequencies, along with converged forces, and displacements, signal stable structures. Additionally, we computed the Raman intensities for all the structures, which can be measured in future experiments.

To calculate the adsorption energies, the values of total energies of H atom ($E_{H_2}$), pristine haeck-BNQD ($E_{haeck-BNQD}$) and all Hatom adsorbed haeck-BNQD models ($E_{model}$) used are given in the Fig 1. The adsorption energy is calculated as follows:

$$\Delta E_{ads} = E_{model} - (E_{haeck-BNQD} + E_H) \qquad (2)$$

The value of adsorption energies are expected to be negative, indicating energy release during the adsorption process, leading to a lowering of the total energy, and stabilizing the system. While performing the calculations on the H-adsorbed systems, the basis-set superposition error (BSSE) arises because the same basis set is shared by both haeck-BNQD and the Hatom. The resultant extra stabilization of intermolecular interaction is taken into account by including the contribution of BSSE into the adsorption energy, leading to the modified formula:

$$\Delta E_{ads} = E_{model} - (E_{haeck-BNQD} + E_H) + E_{BSSE} \qquad (3)$$

Here, $E_{BSSE}$ is calculated using the counterpoise method [55] and is given in Table 2. The order of BSSE is relatively small in the case of B3LYP as compared to wB97XD. The BSSE highly depends on the basis set used. Larger basis sets lead to smaller BSSE, but increase the computational cost.

The catalytic activity of a system is characterized by its Gibbs free energy ($\Delta G$) [56], calculated as:

$$\Delta G = \Delta E_{ads} + 0.24 \qquad (4)$$

where $\Delta E_{ads}$ is the adsorption energy of the system. The best catalytic performance for HER is indicated by $\Delta G = 0$ [14].

Overpotential is also an important parameter to investigate the efficiency of catalyst. Low overpotential implies high catalytic activity, and vice versa. Using the value of Gibbs free energy, we have calculated the overpotential as:

$$\eta = \frac{-|\Delta G|}{e} \qquad (5)$$

## 3. Results and Discussion

For a deep understanding of the catalytic activity of haeck-BNQD and the enhanced reactivity of HER, a detailed study of interactions of H and haeck-BNQD has been performed here. Electronic and structural properties are the main focus of the present study to analyse the effect of haeck-BNQD on the HER mechanism. To effectively determine the active sites of the haeck-BNQD, H is placed at four different positions, discussed in detail the next section. The bond lengths and bond angles of haeck-BNQD are described in the structural properties section. To further investigate the stability of the predicted structures, Raman frequency modes are also discussed in the same section. The rate of reaction will be different because of the different catalytic activity on different active sites. The effects of several factors like PDOS, dipole moment, work-function, and HOMO-LUMO values have also been discussed under the section of electronic properties. It has been seen that the adsorption of H on all four sites is thermodynamically favourable, which we have discussed in detail in Sec.3.3.

## 3.1 Structural Properties

As stated earlier, structural properties have been studied using two different functionals, namely, B3LYP and wB97XD. Geometry optimization of pristine haeck-BNQD for both the functionals was initiated using the geometry parameters reported by Roondhe et al. [57-58] in which the B-N bond lengths vary from 1.39 to 1.48 Å, and for the edge-passivating H atoms, the B-H and N-H bond lengths are 1.20 Å and 1.01 Å, respectively. After relaxation, it is observed that the optimized bond lengths do not change much, and are 1.39-1.48 Å, 1.20 Å, and 1.01 Å, respectively, for the B-N, B-H, and N-H bonds. Two types of bond angles corresponding to N-B-N and B-N-B have been observed, for each of the square and octagon present in the haeck-BNQD. In the following discussion, we have considered the average values because of different B-N-B and N-B-N bond angles in the squares and octagons of haeck-BNQD. The angle of a regular octagon is 135º, but for the optimized geometry, the average value of the N-B-N angle for the octagon is 131.5º which is less than the standard one, and the average value of the B-N-B angle is 138.5º, which is higher. In contrast to the octagon, the N-B-N angle average value for the square is 97.5º, exceeding the standard value of 90.0º, while the average B-N-B angle 82.94º, is lower than the standard value [57,58]. Subsequent to the optimization of pristine haeck-BNQD, the hydrogen adsorption is investigated for possible applications in HER. In order to find the optimized geometries of the H-adsorbed structures, we considered the following four initial configurations for both the functionals with the H atom located (see Fig. 1) at a vertical distance of ~2 Å: (a) on top of boron, (b) on top of nitrogen, (c) on top of the centre of the octagon, and (d) on top of the centre of the square. The final optimized configurations for the adsorbed H atom using the B3LYP functional are (see Fig. 2) : (a) it shifts towards one of the nearest nitrogen atom with the vertical distance of 3.4 Å, (b) it shifts slightly but stays close to the nitrogen itself with the vertical distance of 3.3 Å, (c) it shifts a little to be slightly asymmetrically on top of the octagon, with the vertical distance of 3.7 Å, and (d) it shifts downwards to be on top of the second-nearest N atom, with the vertical distance 3.4 Å. When we use the wB97XD functional, we obtain quite different final configurations for the adsorbed H atom (see Fig. 3): (a) it shifts to the top of octagon, with the vertical distance 4.4 Å, (b) it shifts to the top of the square, with the vertical distance of 4.3 Å, (c) it moves to be on top of a neighbouring octagon, with the vertical distance 4.4, and (d) it moves downwards to be on top of the second-nearest neighbouring N atom, with the vertical distance 4.2 Å.

The formation energy is another parameter to check the effectiveness of haeck-BNQD as a catalyst. The formation energy $E_f$ is the difference between the total energy of the system and the total energy of its constituent particles can be obtained using the formula

$$E_f = \frac{1}{N}(E_{total} - n_B E_B - n_N E_N - n_H E_H - E_H) \qquad (6)$$

where $E_{total}$ is the total energy of the considered system, $E_B, E_H, E_N$ are the energies of the isolated boron, hydrogen, and nitrogen atoms, respectively. Also, $N, n_B, n_N, n_H$ are the total number of atoms in system, and, the total number of boron, nitrogen, and hydrogen atoms, respectively. The formation energies of pristine haeck-BNQD using B3LYP and wB97XD are -6.004 eV and -6.09 eV, respectively. The $E_f$ of H adsorbed haeck-BNQD is presented in Table 2 are calculated using Eq. (6). The negative sign indicates the gain in energy while forming haeck-BNQD from the gaseous states of B, N and H. The lower formation energies of H-adsorbed haeck-BNQDs as compared to their pristine counterpart, show that the adsorption of H is energetically favorable.

To get deeper insights into the vibrational properties of haeck-BNQD along with various hydrogen-adsorbed configurations, we calculated their Raman spectra, presented in Fig. 4. There are a total of 3n-6 vibrational modes, including stretching, twisting, and bending deformations, where n represents the total number of atoms in the structure in hydrogen adsorbed systems. Here, we have 49 atoms in the hydrogen-adsorbed haeck-BNQDs, therefore, there will be (3*49)-6 = 141, vibrational modes for each structure. All the observed modes have real frequencies confirming the stability of the structures during HER. It is clear from Fig. 4(a) that there are three main frequency regions for both the B3LYP and wB97XD functionals: (a) low-energy region with frequencies 0-1600 $cm^{-1}$, (b) middle-energy region with frequency range 2500-2750 $cm^{-1}$, and (c) the high-energy region with frequencies in the range 3500-3750 $cm^{-1}$. The first region corresponds to B-N deformations of the QD surface in the 0-500 $cm^{-1}$ range, with the adsorbed H bending, twisting, and stretching. The region between 500-1500 $cm^{-1}$ again has low intensity, and corresponds to B-N stretching modes. The region with frequencies in the range 2500-2750 $cm^{-1}$ shows high-intensity peaks because of B-H stretching vibrations, with the highest intensity peak at 2580 $cm^{-1}$ frequency. The high-frequency region of 3500-3750 $cm^{-1}$ represents N-H stretching vibrations, and exhibits the highest intensity peak among all the regions located at 3570 $cm^{-1}$. In case of B3LYP functional, the N-H stretching modes show highest intensity peak, whereas in case of wB97XD functional B-H stretching modes exhibit the highest intensity peaks.

## 3.2 Electronic Properties

To obtain deeper insights into the influence of the electrocatalytic activity on the HER performance, we have studied the electronic properties such as orbital energies of the highest occupied molecular orbital (HOMO) and the lowest unoccupied molecular orbital (LUMO), dipole moment ($p$), work function ($\phi$), total density of states (TDOS) along with partial density of states (PDOS), charge transfer, and the molecular electrostatic potential (MESP) for all the considered structures. In the following, the HOMO and LUMO orbital energies are abbreviated as $E_{HOMO}$ and $E_{LUMO}$, respectively. The $E_{HOMO}$ and $E_{LUMO}$ values using the B3LYP functional of H-adsorbed structures are nearly equal to ~-6.6 and ~-1.46 eV, respectively, whereas for wB97XD functional the values are ~-8.7 and ~0.94 eV. We also note that with the hydrogen adsorption, $E_{HOMO}$ values for all the four configurations are slightly reduced with respect to the pristine haeck-BNQD (see Table1), depicting their reduced electron-donor ability. Similarly, reduced $E_{LUMO}$ values depict increased electron-acceptor abilities of the hydrogen-adsorbed BNQD systems using both the functionals.

The dipole moments in Table 1 represent the global molecular reactivity indicating the net molecular polarity and anisotropy present in the pristine and Hadsorbed haeck-BNQD structures. In B3LYP, the dipole moments range from 0.065 to 0.123D. The dipole moment for the structure with the H-adsorbed on top of the square is 0.065 D, which is smaller as compared to other three structures. However, for the wB97XD functional, the dipole moments changes from 0.624D (pristine haeck-BNQD) to 0.533D (smallest in case of H-O-haeck-BN).

The work function, which is an important property of a system related to its electronic structure, is defined as the average of the ionization potential (IP) and the electron affinity (EA):

$$\phi = \frac{IP+EA}{2} \quad (7)$$

The work function values for nanostructures such as the ones considered here, broadly indicate the energy required to eject an electron from the system. From Table 1, we note that the work function values present significant changes in B3LYP functional, however, they do not display much variations for the wB97XD functional calculations on various structures.

MESP yields information about the reactivity of electrophilic and nucleophilic species from the charge density distribution. In MESP plots presented in Figs. 5 and 6, the blue and the red colours indicate the affinity of nucleophilic and electrophilic species, respectively. From Figures 5 and 6, it is obvious that the haeck-BNQD has blue colour on its boundary, but the structure with Hadsorbed on the top of the nitrogen (Fig. 6(b)) has red colour, exhibiting the electrophilic nature of the nitrogen because of its high electronegativity.

In Fig. 7(a-b), we present the plots of HOMO and LUMO orbitals of all the structures computed using both the exchange-correlation functionals. From the figure it is obvious thatLUMO orbitals of all the configurations are similar to each other for each functional. Similarly, the same pattern holds for all the HOMO orbitals in B3LYP functional except for N-haeck-BNQD, for which the HOMO orbital computed by both the functionals are different, with the charge density concentrated near the centre of the structure.

TDOS provides information about available single-particle states both below and above the Fermi level. In Fig. 8, we present the TDOS calculated for all the four hydrogen-adsorbed structures along with the pristine haeck-BN, and we observe the following trends: (a) for a given functional, the TDOS plots for all the structures are very similar to each other, with coincident peaks (b) in the energy range below the Fermi level, for all the structures TDOS has four peaks irrespective of the functional used, (c) with the B3LYP functional, for energies above the Fermi level, all structures display two plateaus in their TDOS, and (d) above the Fermi level TDOS computed using the wB97XD functional exhibit single plateaus for the all the hydrogen-adsorbed structures. We note that the locations of TDOS plateaus in the unoccupied energy range also overlap for all the four structures, for a given exchange-correlation functional.Another useful quantity related to the electronic structure is the partial density of states (PDOS) which describes the individual contributions of the constituent atoms to the TDOS, both for the occupied and unoccupied states. We computed the PDOS using the computer package Multiwfn [47], and the results are shown in Fig 8. For both the functionals, HOMO has a contribution from all the atoms B, N, and H, but the N atom contributes the most, followed by B and H atoms. Unlike HOMO, major contributions to LUMO are made by B and H. On the other hand, for the infinite haeckelite BN nanosheets, it has been reported that the $2p_z$ orbitals of both B and N atoms contribute to the PDOS of both valence and conduction bands [59].

The atomic charges redistribute on haeck-BNQD after adsorption of the H atom, therefore, we computed the Mulliken charges of all the atoms in the system, and the results are presented in Fig. 9. The labelling in the haeck-BNQD is such that the first 16 atoms are hydrogen, 17-32 are boron, and 33-48 are nitrogen atoms, and lastly, the label for the adsorbed H atom is 49. In both of the functionals, we have seen inhomogeneous positive charges on boron and negative charges on nitrogen. Both positive and negative charges on H are present depending upon the nature of the bonded atom. As we know, the N atom is more electronegative than H, results in a positive charge on H, and B is less electronegative than H, resulting in the negative charge on H. When we add H to the haeck-BNQD using both the functionals, H has positive charge labelled by H49 in the figure. The maximum values of positive and negative charges on (B, N) for B3LYP are (0.516, -0.641), and (0.536, -0.689) for wB97XD. From the quantitative analysis of the Mulliken charges, it is clear that these charges are directly related to the nature of the chemical bonds, and, therefore, affect various other properties such as dipole moment, polarizability, electronic structure, etc.

## 3.3 HER

In the previous sections we discussed the electronic and structural properties of haeck-BNQD structures relevant for their catalytic activity. Next, we discuss the parameters such as adsorption energy ($\Delta E_{ads}$), Gibbs free energy ($\Delta G$), overpotential ($\eta$) that are relevant to the performance of haeck-BNQD in HER. In Table 2, we present the calculated values of all these quantities. It is obvious that the lower values of adsorption energies ($\Delta E_{ads}$) indicate better HER activity. For the B3LYP functional, the least value of adsorption energy -0.024 eV is observed for adsorption on top of boron and nitrogen atoms, while with the wB97XD functional the lowest value of adsorption energy -0.008 eV is on top of the centre of the square.

Gibbs free energy $\Delta G$ is one of the key parameters that decide the efficiency of catalytic activity of HER, as discussed next. To investigate the HER performance of the haeck-BNQD, we have employed the Sabatier principle [60]. To calculate the HER activities of the haeck-BNQD system, the interaction between H and the system was studied by computing the optimized geometries of the H atom adsorption. The water reduction occurs on cathode, which, together with the electrolyte, serves as a catalyst in the acidic media for HER. The HER can proceed by two possible reaction pathways namely Volmer−Heyrovsky and Volmer−Tafel mechanisms [14], for each of which the adsorption energy of hydrogen atom is an important parameter in deciding the surface catalytic activity. According to Sabatier principle [61], $\Delta G \approx 0$ implies ideal HER thermodynamics [62]. Physically, this means that in order to achieve low overpotential for HER, the H adsorption should neither be too strong ($\Delta G \ll 0$), nor too weak ($\Delta G > 0$). This is because if the $\Delta G$ value is too small, H is so strongly attached to the surface that it will be hard to desorb it from the haeck-BNQD, while if it is too large, H atom will not get adsorbed on the quantum dot at all. Thus, the ideal value for $\Delta G$ is zero for an efficient HER. In order to examine the HER activities of the considered systems, we have calculated $\Delta G$ from the $E_{ads}$ values of H on the haeck-BNQD using Eq. 4 for all the four hydrogen adsorption configurations, and the results are listed in Table 2. From the B3LYP functional-based calculations it is obvious that $\Delta G$ is nearly equal to ~0.21 eV for all the considered sites. However, calculations performed using the wB97XD functional indicate $\Delta G$ values of 0.23 eV for all the configurations. Thus we conclude that the low values of $\Delta G$ predicted by both the functionals imply that HER is favourable for these systems. This point is further reinforced in Fig. 10, in which Gibbs free energy is plotted as a function of reaction coordinates.

The overpotentials have been calculated using the Eq. (5), and are also presented in Table 2. From a physical point of view, the overpotential ($\eta$) is the extra potential required to drive the electrochemical water-splitting reaction, implying that the lesser its value, the higher the electrocatalytic activity. From the calculated values of overpotential presented in Table 2 it is obvious thatfor all the H-adsorbed configurations, and both the exchange-correlation functionals, the values are rather small, lying in the range 0.216 V –0.234 V. Thus, we conclude that haeck-BNQD reduces the overpotential considerably, and enables a highly efficient HER. Also, the minimum value of overpotential 0.216 V was obtained from the B3LYP functional-based calculations for two adsorption configurations: (a) on top of the boron, and (b) on the top of the nitrogen.

## 4. Conclusion

In the present study, a new haeckelite structure of boron nitride quantum dots has been predicted. The absence of imaginary frequencies, along with converged forces and displacements during the optimization process suggest that the structure is completely stable. For the haeck-BNQD structure, we get the most favourable values of hydrogen adsorption and Gibbs free energy using the wB97XD functional, which are roughly the same for all the considered adsorption configurations. This, along with their similar overpotential values, suggests strong enhancement of the catalytic activity of these structures as far HER is concerned. Therefore, based upon the results of the present study, we hope that experimentalists will consider the use of haeck-BNQDas an electrocatalyst for the hydrogen evolution reaction.

## Acknowledgements

One of the authors, RJ acknowledges financial assistance from CSIR JRF. VS acknowledges the support through the Institute Post-Doctoral Fellowship (IPDF) of Indian Institute of Technology Bombay.

**Table 1:** Calculated HOMO/LUMO energies, dipole moments, and work functions of all the considered systems, computed using both B3LYP and wB97XD functionals.

| Functional | Haeck-BNQD Systems | $E_{HOMO}$ (eV) | $E_{LUMO}$ (eV) | $E_f$ (eV) | Dipole moment p (D) | $\phi$ (eV) |
|---|---|---|---|---|---|---|
| B3LYP | Pristine haeck-BNQD | -6.60 | -0.978 | -6.0048 | 0.213 | 3.791 |
| | H-B-haeck-BNQD | -6.604 | -1.457 | -5.8828 | 0.102 | 4.03 |
| | H-N-haeck-BNQD | -6.604 | -1.459 | -5.8827 | 0.111 | 4.031 |
| | H-O-haeck-BNQD | -6.603 | -1.388 | -5.8827 | 0.123 | 3.995 |
| | H-S-haeck-BNQD | -6.604 | -1.434 | -5.8827 | 0.065 | 4.019 |
| wB97XD | Pristine haeck-BNQD | -8.698 | 0.943 | -6.090 | 0.624 | 3.877 |
| | H-B-haeck-BNQD | -8.669 | 0.943 | -5.966 | 0.534 | 3.863 |
| | H-N-haeck-BNQD | -8.698 | 0.944 | -5.966 | 0.598 | 3.877 |
| | H-O-haeck-BNQD | -8.669 | 0.943 | -5.966 | 0.533 | 3.863 |
| | H-S-haeck-BNQD | -8.698 | 0.944 | -5.966 | 0.599 | 3.877 |

**Table 2:** Calculated adsorption energy ($E_{ads}$), Gibbs free energy ($\Delta G$), and overpotential ($\eta$) of all considered systems.

| Functional | Haeck-BN Systems | $\Delta E_{ads}$ (eV) | $\Delta E_{ads}^{BSSE}$ (eV) | $\Delta G$ (eV) | overpotential $\eta$ (V) | Theoretical (Others) $\Delta E_{ads}$ (eV) |
|---|---|---|---|---|---|---|
| B3LYP | H-B-haeck-BNQD | -0.024 | -0.024 | 0.216 | 0.216 | |
| | H-N-haeck-BNQD | -0.024 | -0.024 | 0.216 | 0.216 | |
| | H-O-haeck-BNQD | -0.023 | -0.023 | 0.217 | 0.217 | -2.64 eV[36] -0.031 eV[64] 0.60 eV[65] 1.930 eV[66] 1.57 eV[63] |
| | H-S-haeck-BNQD | -0.023 | -0.023 | 0.217 | 0.217 | |
| wB97XD | H-B-haeck-BNQD | -0.0064 | -0.0063 | 0.234 | 0.234 | |
| | H-N-haeck-BNQD | -0.0072 | -0.0071 | 0.233 | 0.233 | |
| | H-O-haeck-BNQD | -0.0064 | -0.0063 | 0.234 | 0.234 | |
| | H-S-haeck-BNQD | -0.008 | -0.008 | 0.232 | 0.232 | |

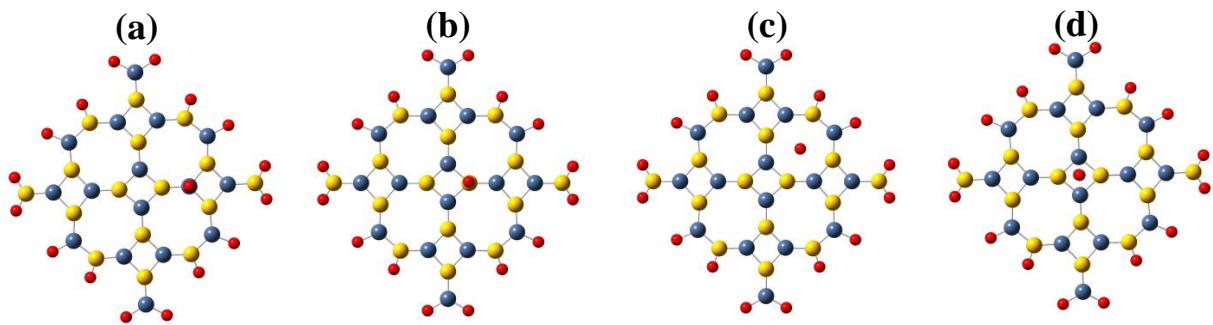

**Figure 1:** Initial structures of hydrogen over (a) B-haeck-BNQD (b) N-haeck-BNQD (c) O-haeck-BNQD and (d) S-haeck-BNQD. Boron, Nitrogen and Hydrogen are blue, yellow and red respectively.

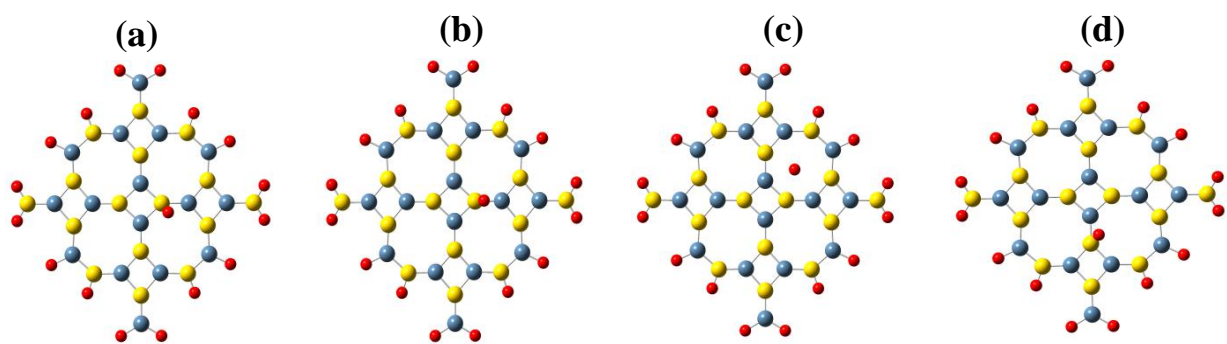

**Figure 2:** Optimized structures of hydrogen over (a) B-haeck-BNQD (b) N-haeck-BNQD (c) O-haeck-BNQD and (d) S-haeck-BNQD with B3LYP functional. Boron, Nitrogen and Hydrogen are blue, yellow and red respectively.

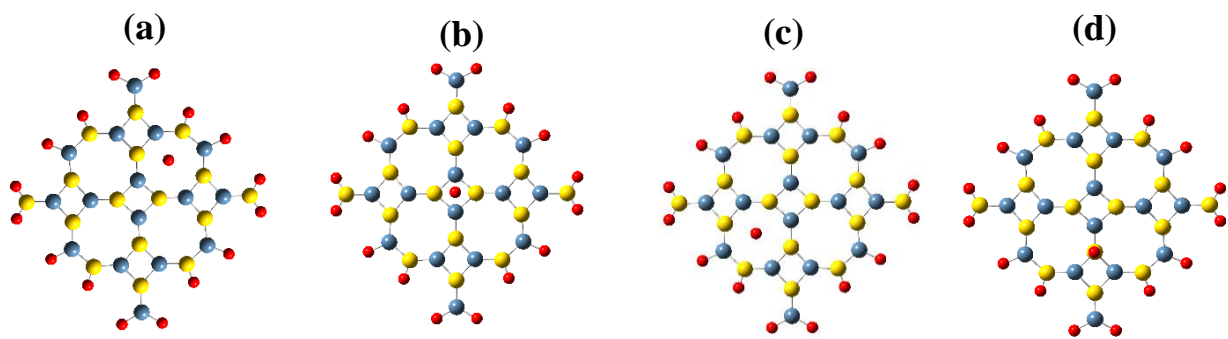

**Figure 3:** Optimized structures of hydrogen over (a) B-haeck-BNQD (b) N-haeck-BNQD (c) O-haeck-BNQD and (d) S-haeck-BNQD with wB97XD functional. Boron, Nitrogen and Hydrogen are blue, yellow and red respectively

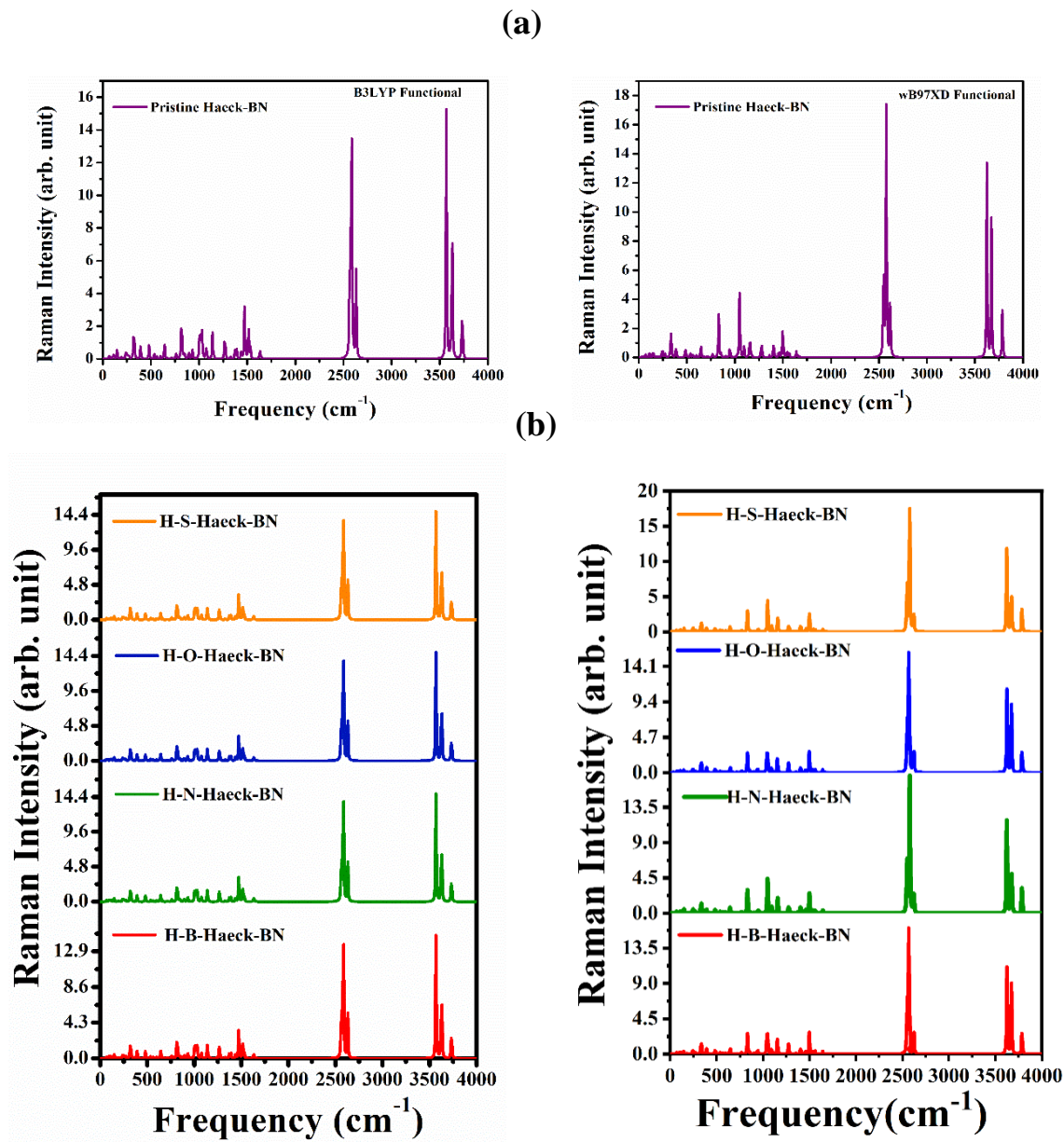

**Figure 4:** Raman plots of (a) pristine haeck-BNQD (b) hydrogen over haeck-BNQD.

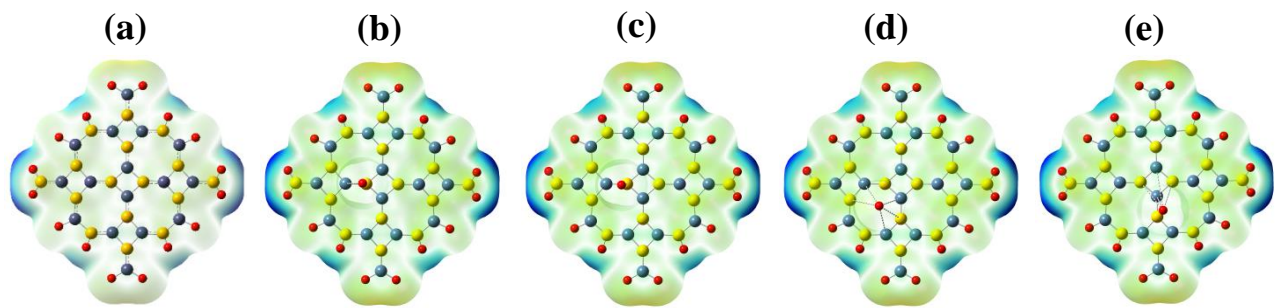

**Figure 5:** MESP plots for (a) pristine haeck-BNQD, and H-adsorbed systems (b) B-haeck-BNQD (c) N-haeck-BNQD (d) O-haeck-BNQD and (e) S-haeck-BNQD, all computed using the B3LYP functional.

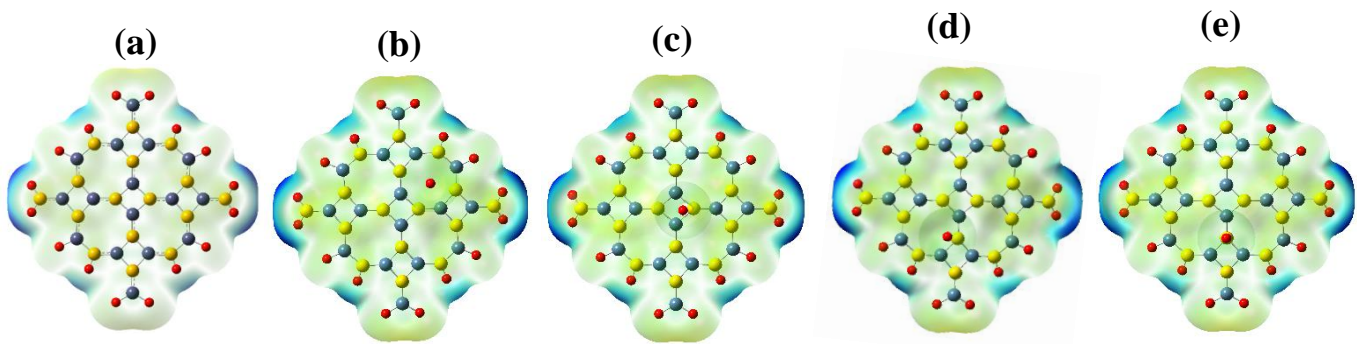

**Figure 6:** MESP plots for (a) pristine haeck-BNQD, and H-adsorbed systems (b) B-haeck-BNQD (c) N-haeck-BNQD (d) O-haeck-BNQD and (e) S-haeck-BNQD, all computed using the wB97XD functional.

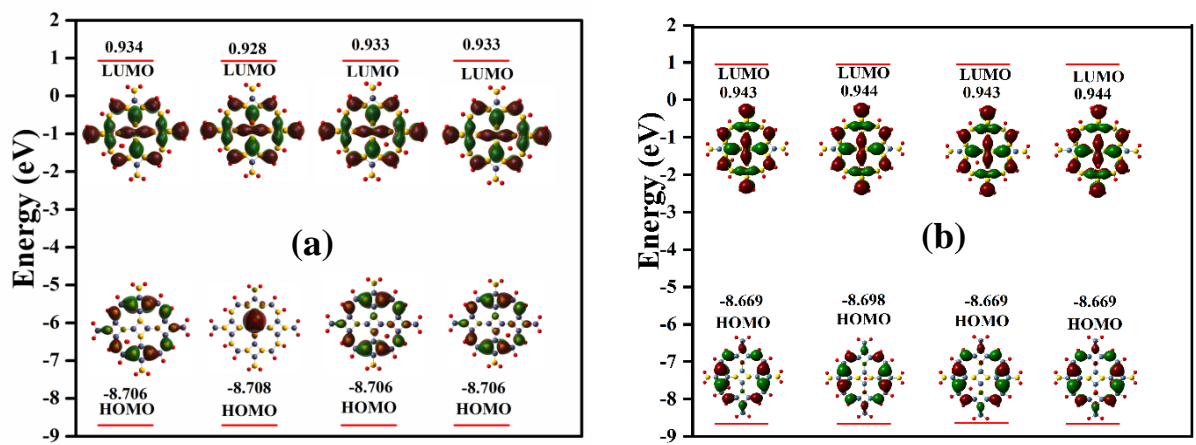

**Figure 7:** HOMO-LUMO plots of various H-adsorbed configurations computed using the (a) B3LYP, and (b) wB97XD functionals.

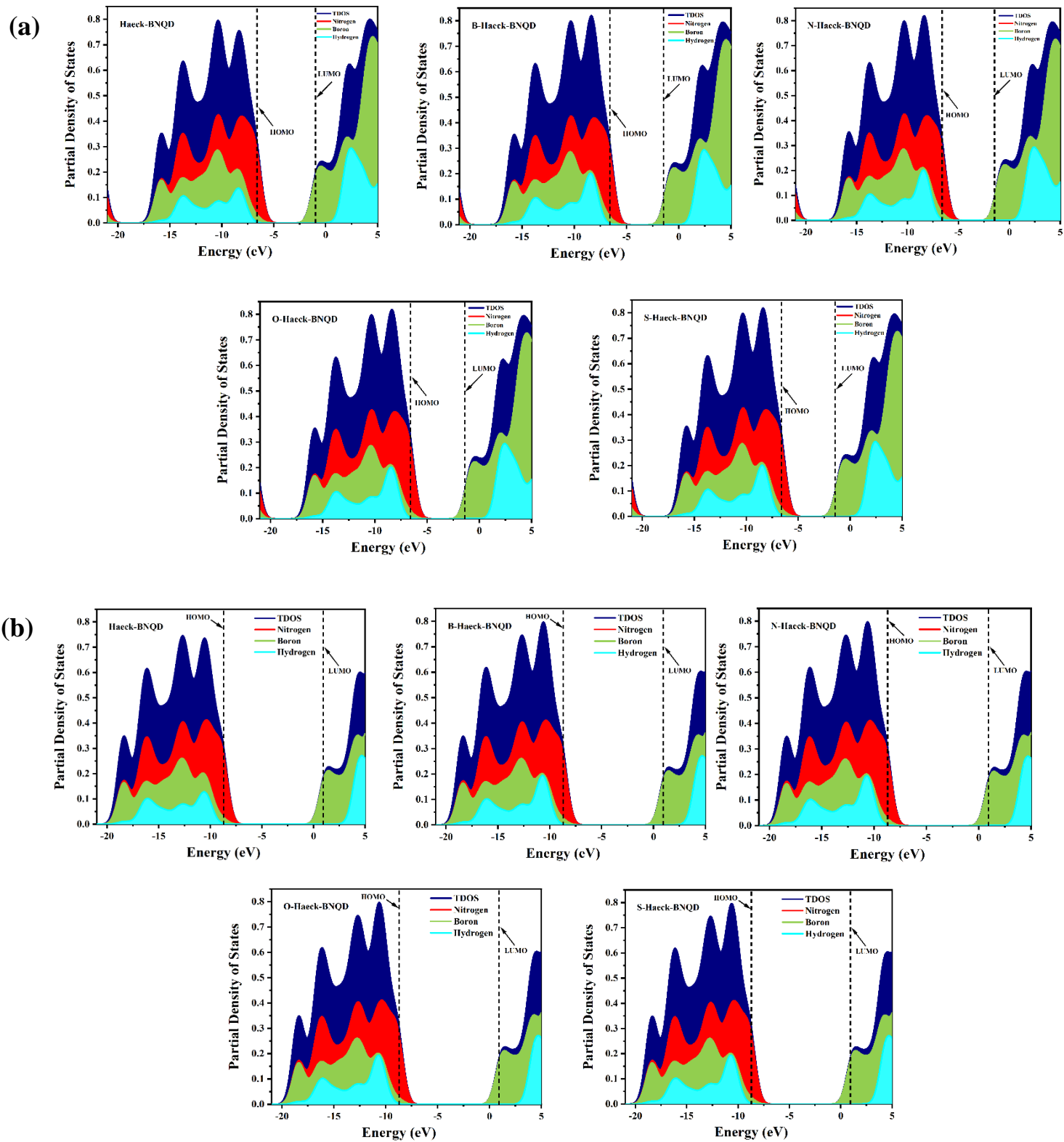

**Figure 8:** Total and partial density of states plots of various H-adsorbed systems computed using the (a) B3LYP, and (b) wB97XD functionals.

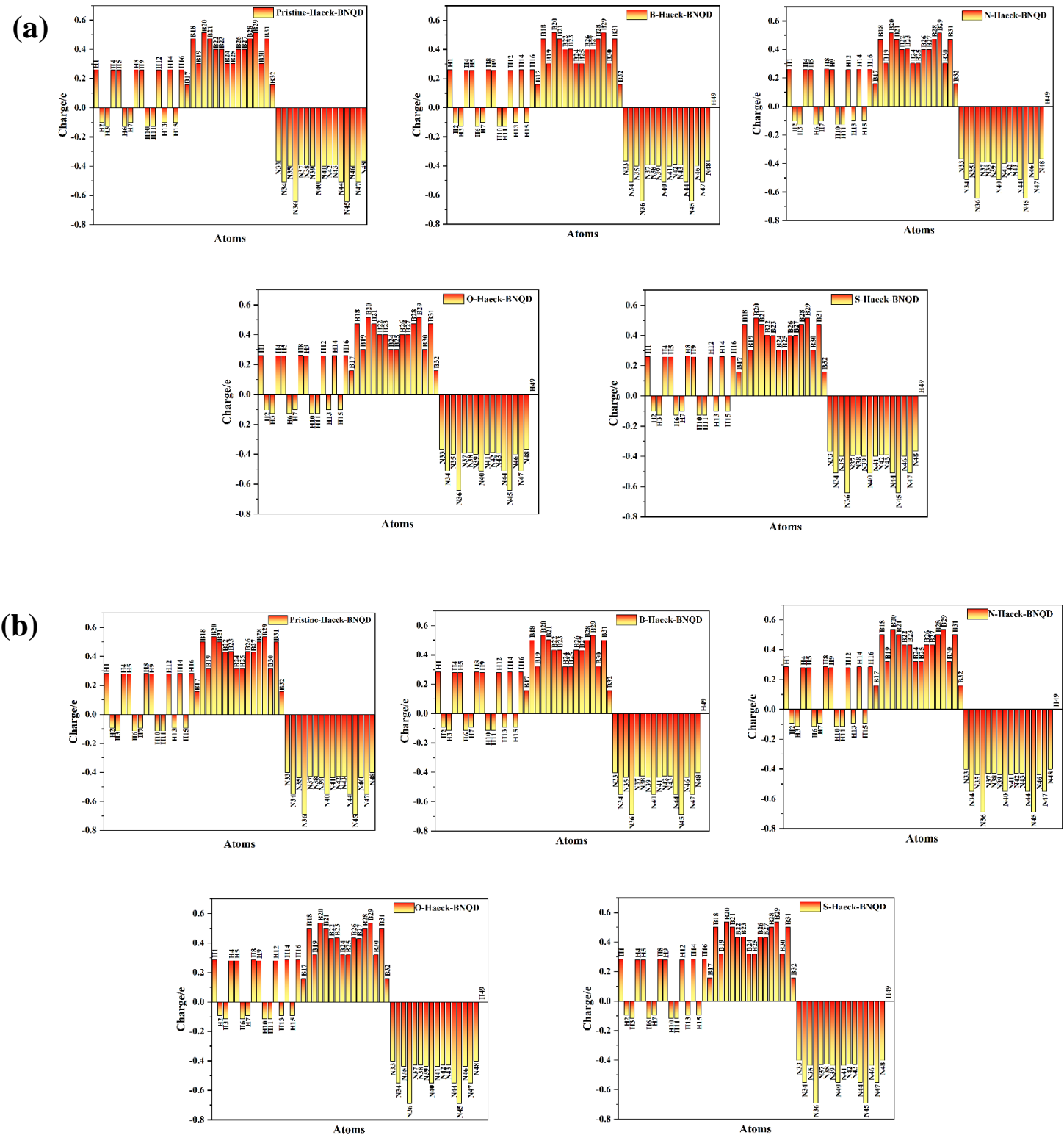

**Figure 9:** Mulliken charge transfer plots of H-adsorbed systems computed using the (a) B3LYP, and (b) wB97XD functionals

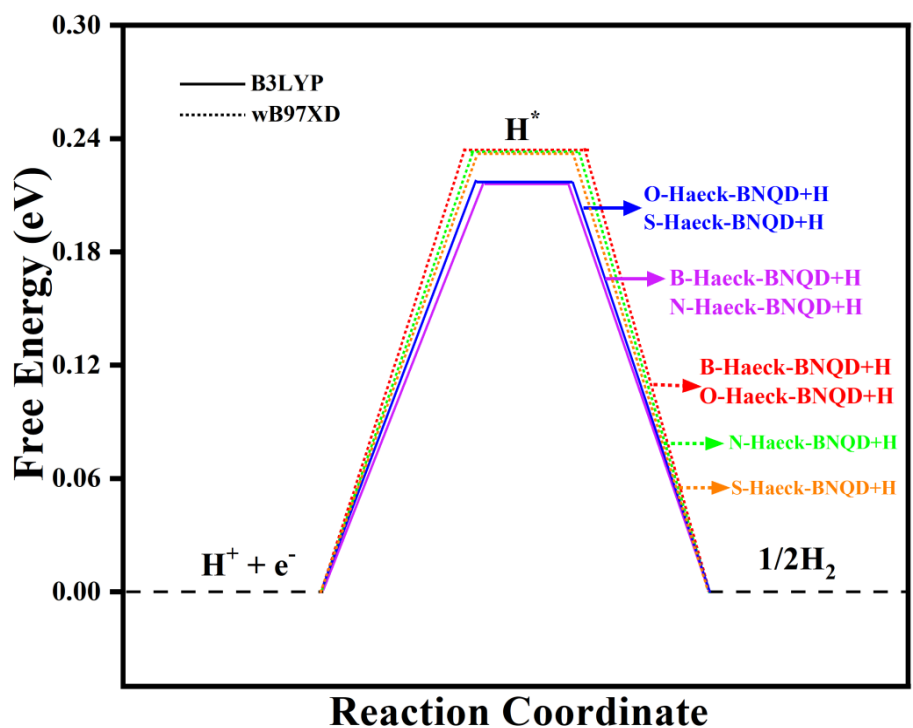

**Figure 10:** The free energy diagrams for various H-adsorbed haeck-BNQD structures. The solid and dashed lines represent Gibbs free energies computed using the B3LYP and wB97XD functionals, respectively.